\documentclass[sigconf]{acmart}

\usepackage{bm}
\usepackage{subfig}
\usepackage{natbib}
\usepackage[utf8]{inputenc} % allow utf-8 input
\usepackage[T1]{fontenc}    % use 8-bit T1 fonts
\usepackage{booktabs}       % professional-quality tables
\usepackage{amsfonts}       % blackboard math symbols
\usepackage{nicefrac}       % compact symbols for 1/2, etc.
\usepackage{microtype}      % microtypography
\usepackage{caption}
\usepackage{url}            % simple URL typesetting
\usepackage{multirow}
\usepackage{enumitem}
\usepackage[ruled,vlined]{algorithm2e}

\usepackage{pifont}
\usepackage{bbding}
\usepackage{tikz,xcolor}% Make Orcid icon
\usepackage{tcolorbox}
\usepackage{wrapfig}
\usepackage{listings}
\usepackage[normalem]{ulem}
\useunder{\uline}{\ul}{}

\newcommand{\ie}{\textit{i.e.}}
\newcommand{\eg}{\textit{e.g.}}
\newcommand{\wrt}{\textit{w.r.t.}}

\theoremstyle{plain}
\newtheorem{theorem}{Theorem}[section]

\theoremstyle{definition}

\theoremstyle{remark}
\newtheorem{remark}[theorem]{Remark}

\AtBeginDocument{%
  }

\setcopyright{acmlicensed}
\copyrightyear{2018}
\acmYear{2018}
\acmDOI{XXXXXXX.XXXXXXX}
\acmConference[Conference acronym 'XX]{Make sure to enter the correct
  conference title from your rights confirmation email}{June 03--05,
  2018}{Woodstock, NY}
\acmISBN{978-1-4503-XXXX-X/2018/06}

\begin{document}

%%
%% The "title" command has an optional parameter,
%% allowing the author to define a "short title" to be used in page headers.
\title{Markovian Pre-Trained Transformer for Next-Item Recommendation}

\author{Cong Xu}
\affiliation{%
  \institution{East China Normal University}
  \city{Shanghai}
  \country{China}
}
\email{congxueric@gmail.com}

\author{Guoliang Li}
\affiliation{%
  \institution{Tsinghua University}
  \city{Beijing}
  \country{China}
}
\email{liguoliang@tsinghua.edu.cn}

\author{Jun Wang}
\affiliation{%
  \institution{East China Normal University}
  \city{Shanghai}
  \country{China}
}
\email{wongjun@gmail.com}

\author{Wei Zhang}
\affiliation{%
  \institution{East China Normal University \\ \& Shanghai Innovation Institute}
  \city{Shanghai}
  \country{China}
}
\email{zhangwei.thu2011@gmail.com}

%%
%% By default, the full list of authors will be used in the page
%% headers. Often, this list is too long, and will overlap
%% other information printed in the page headers. This command allows
%% the author to define a more concise list
%% of authors' names for this purpose.
\renewcommand{\shortauthors}{Xu et al.}

%%
%% The abstract is a short summary of the work to be presented in the
%% article.
\begin{abstract}
We introduce the Markovian Pre-trained Transformer (MPT) for next-item recommendation,
a transferable model fully pre-trained on \textit{synthetic} Markov chains, 
yet capable of achieving state-of-the-art performance by fine-tuning a lightweight adaptor.
This counterintuitive success stems from the observation of the `Markovian' nature:
advanced sequential recommenders coincidentally rely on the latest interaction to make predictions, 
while the historical interactions serve mainly as auxiliary cues for inferring the user's \textit{general, non-sequential} identity.
This characteristic necessitates the capabilities of a universal recommendation model to \ding{182} effectively summarize the user sequence,
with \ding{183} particular emphasis on the latest interaction.
MPT inherently has the potential to be universal and transferable.
On the one hand, when trained to predict the next state of Markov chains, 
it acquires the capabilities to \ding{182} estimate transition probabilities from the context (one adaptive manner for summarizing sequences)
and \ding{183} attend to the last state to ensure accurate state transitions.
On the other hand, unlike the heterogeneous interaction data, 
an unlimited amount of controllable Markov chains is available to boost the model capacity.
We conduct extensive experiments on five public datasets from three distinct platforms to 
validate the superiority of Markovian pre-training over traditional recommendation pre-training and recent language pre-training paradigms.
Our code is available at \url{https://github.com/BDML-lab/MPT}.
\end{abstract}


% \ccsdesc[500]{Computer systems organization~Neural networks}

%%
%% Keywords. The author(s) should pick words that accurately describe
%% the work being presented. Separate the keywords with commas.

\received{20 February 2007}
\received[revised]{12 March 2009}
\received[accepted]{5 June 2009}

%%
%% This command processes the author and affiliation and title
%% information and builds the first part of the formatted document.
\maketitle

\section{Introduction}

The pre-training and fine-tuning workflow has become the \textit{de facto} paradigm across diverse domains~\cite{chen2020simclr,devlin2019bert,radford2021clip},
allowing for significant computational savings through the efficient knowledge transfer.
In next-item recommendation, 
a model trained over users' historical interactions is unfortunately domain-specific even when leveraging text embeddings as input.
Consequently tedious retraining must be performed over and over to keep adaptive to the latest scenarios.
Although considerable efforts~\cite{ding2021zesrec,hou2022unisrec,li2023recformer} have been devoted to developing a universal pre-trained recommender, 
the performance still lags far behind those trained from scratch. 
This discrepancy could be attributed to the heterogeneity of public datasets, 
ranging from diverse user behaviors to non-negligible domain gaps~\cite{ju2025revisiting,ye2025beyond}.
Some researchers have therefore turned to foundation models 
that are not specifically pre-trained for recommendation tasks, \eg, Large Language Models (LLMs)~\cite{li2023e4srec,liao2024llara};
however, these approaches remain inadequate in terms of accuracy and, more critically, efficiency.
In this paper, we aim to pre-train a universal Transformer-based~\cite{vaswani2017attention} model for next-item recommendation by addressing the following two fundamental questions:  
\textbf{(RQI)} Which capabilities are transferable across diverse recommendation scenarios?
\textbf{(RQII)} What types of data are sufficient to stimulate the required capabilities?

\begin{figure*}[t]
	\centering

  \subfloat[Model \& Dataset]{
    \includegraphics[width=0.665\textwidth]{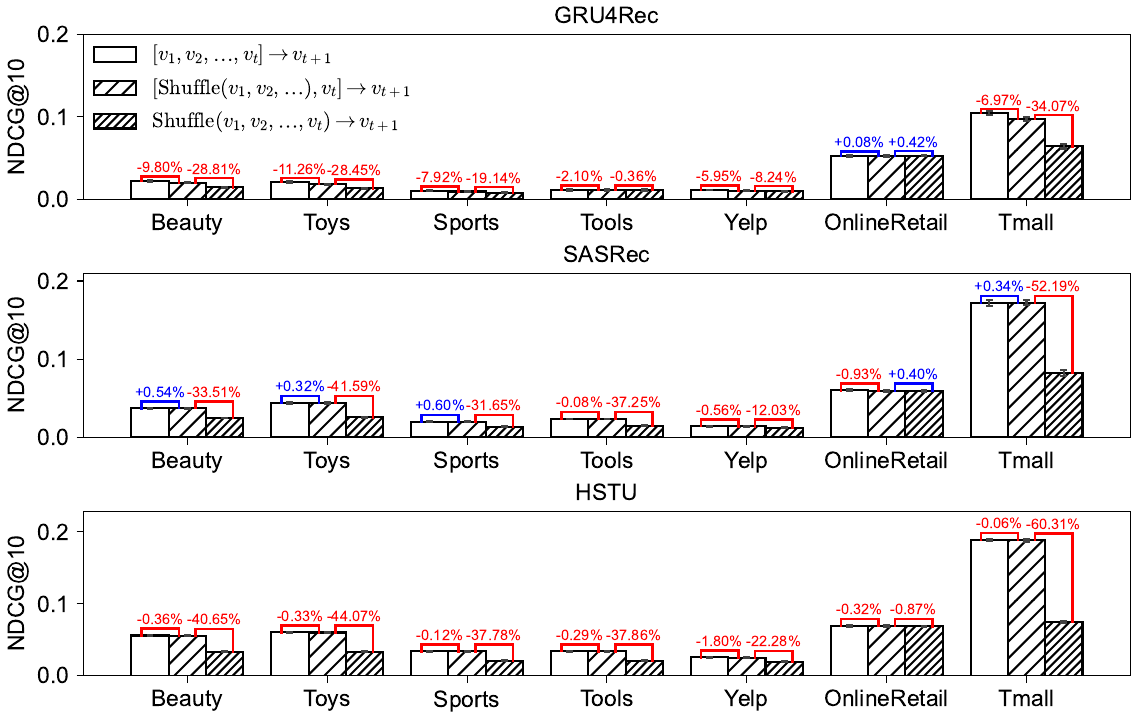}
    \label{fig-order-model-dataset}
  }
  \subfloat[Filter \& Loss \& Size]{
    \includegraphics[width=0.315\textwidth]{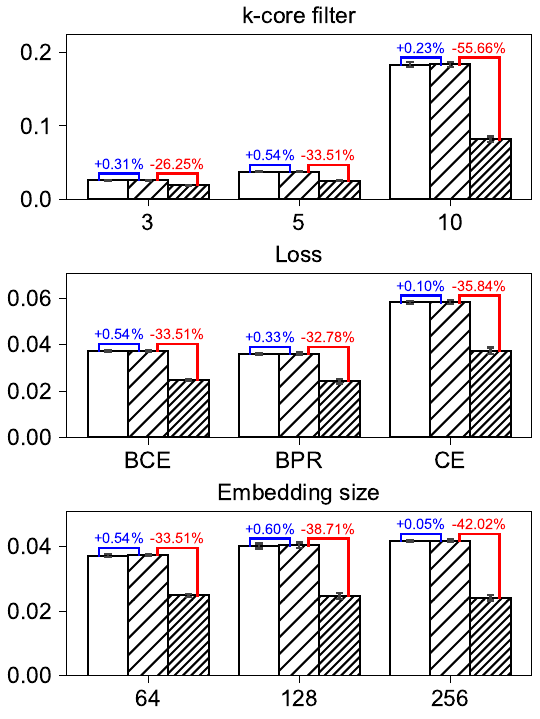}
    \label{fig-order-ablation}
  }
	\caption{
      Comparisons of the impact of sequence ordering (during inference).
      $[v_1, v_2, \ldots, v_t]$: chronologically ordered sequence;  
      $[\text{Shuffle}(v_1, v_2, \ldots), v_t]$: partially shuffled sequence;  
      $\text{Shuffle}(v_1, v_2, \ldots, v_t)$: completely shuffled sequence.
      \textbf{(a)} Performance across various models and datasets.
      \textbf{(b)} Performance across filtering strategies, training objectives, and embedding sizes.
	}
	\label{fig-order}
\end{figure*}

We first empirically identify the `Markovian' characteristics of next-item recommendation
through a systematic evaluation of sequential recommenders, including GRU4Rec~\cite{hidasi2015gru4rec}, SASRec~\cite{kang2018sasrec}, and HSTU~\cite{zhai2024hstu}, across several public datasets.
Given a chronologically ordered sequence of historical interactions $[v_1, v_2, \ldots, v_t]$, 
we compare in \figurename~\ref{fig-order} how NDCG varies when the input sequences 
are either partially shuffled (with the latest interaction $v_t$ fixed) or completely shuffled.
If the chronological order is essential for a sequential model to perform complex yet more accurate inference (\eg, involving logical reasoning), 
both the shuffled variants would be expected to cause substantial performance degradation.
Surprisingly, our results show that, as long as the latest interaction remains fixed, 
no statistically significant differences would be observed across datasets or platforms.
These observations suggest the current advanced sequential recommenders make next-item predictions in a Markovian fashion: they are capable of \ding{182} summarizing non-sequential user preferences from historical interactions, 
while \ding{183} placing particular emphasis on the most recently interacted item.
Importantly, these capabilities are transferable across diverse datasets and platforms, 
pointing to a promising direction for developing specialized pre-trained recommendation models.

Analogous to the fundamental role of mathematical data in stimulating the logical reasoning ability of LLMs~\cite{huan2025does,liu2024deepseek},
it is worthwhile to ask whether specific types of data can similarly stimulate the required recommendation capabilities.
The raw interaction data, throughly mined by previous studies already, might not be clean and controllable enough~\cite{ju2025revisiting,ye2025beyond} for this purpose.
We thus resort to synthetic data.
Recently,
Lepage et al.~\citep{lepage2025markov} demonstrated that a Transformer model
can learn to estimate the transition probabilities of a Markov chain from context instead of memorizing the training patterns.
Concretely,
the Markovian Pre-trained Transformer (MPT), fully pre-trained on synthetic Markov chains, 
can accurately predict the next possible state of an arbitrary trajectory,
even if the sampled trajectory has never been encountered during training.
This indicates that MPT must possess the capabilities to \ding{182} estimate intrinsic transition probabilities from the context (one adaptive manner for summarizing sequences) and \ding{183} attend to the last state to ensure accurate state transitions.
They correspond precisely to the aforementioned recommendation capabilities, one by one, 
also suggesting a novel Markovian pre-training and recommendation fine-tuning paradigm. 
Moreover,
unlike the heterogeneous interaction data, 
the synthetic Markov chains are both unlimited and controllable.

Recommendation fine-tuning on top of MPT can be readily performed using only a lightweight input adaptor, 
or alternatively augmented with parameter-efficient fine-tuning strategies such as LoRA~\cite{hu2022lora}.
We thoroughly investigate the superiority of MPT across five public datasets from three distinct platforms. 
In most scenarios, state-of-the-art performance can be achieved using the input adaptor alone, yielding relative improvements of approximately 8\% over traditional recommenders and 25\% over pre-trained methods.
Notably, such improvements are progressively realized after pre-training on 10B tokens.
By contast, collecting such a large volume of \textit{clean} and \textit{controllable} real-world data is inherently challenging.
In addition,
by comparing attention patterns, 
the effectiveness of Markovian pre-training over recommendation pre-training and language pre-training can be further elucidated.

Our contributions to the community are three-fold:
\begin{itemize}[leftmargin=*]
\item We systematically identify two transferable capabilities for next-item recommendation, 
rather than limiting our analysis to obscure concepts like short-term or long-term interests. 
These findings may inspire the research community to develop more specialized architectures and optimization strategies.

\item While prior work on recommendation pre-training primarily focuses on leveraging heterogeneous interaction data or resorting to language pre-training, the success of MPT highlights a fundamentally novel paradigm: exploiting synthetic data to endow recommenders with the required capabilities. 
We believe that a focus on capability stimulation is central to effective recommendation pre-training.

\item We also examine the data and model scaling behaviors of MPT. Notably, 
unlike the well-known scaling laws for LLMs~\cite{kaplan2020scaling, hoffmann2022scaling}, 
Markovian pre-training converges to a theoretical limit governed by the Bayes estimator~\cite{berger2013statistical}.
Such a theoretical limit may further offer certain guidance for recommendation scaling~\cite{zhang2024wukong}.

\end{itemize}

\section{Markovian Nature of Next-Item Prediction}
\label{section-markovian-nature}

Given a user's chronological interaction sequence $[v_1, v_2, \ldots, v_t]$,
the next-item recommendation task aims to predict the subsequent item $\hat{v}_{t+1} \in \mathcal{V}$ that the user is likely to engage with.
As expected, 
architectures such as RNNs~\cite{hidasi2015gru4rec} and Transformers~\cite{kang2018sasrec} have demonstrated superior performance in this field
owing to their strong capability to capture sequential dependencies.
They typically encode the interaction sequence into a vector representation
and compute relevance scores over the item set $\mathcal{V}$ using inner product or cosine similarity.
The top-ranked items are then considered as candidates for recommendation to the user.
To facilitate the examination of the `Markovian' nature of next-item recommendation, 
we formulate the prediction task as a process of maximizing the conditional probability:
\begin{align}
  \hat{v}_{t+1} = \underset{v \in \mathcal{V}}{\text{argmax}} \:
  \mathbb{P} \left(
    v \,|\, v_t, v_{t-1}, \ldots, v_1; \Theta
  \right),
\end{align}
where $\mathbb{P}(\cdot \mid \cdot; \Theta)$ 
gives the probability estimated by a sequential model conditioned on the interaction sequence
and $\Theta$ represents its learnt parameters.
For instance, GRU4Rec recurrently incorporates items;
SASRec captures sequential dependencies in parallel via causal attention; 
and HSTU further exploits pointwise aggregated attention and time-based positional encoding.
As increasingly powerful architectures are employed for next-item recommendation, 
an important question arises: 
do they perform next-item prediction in an increasingly complex fashion?

In this section, 
we are to demonstrate that sophisticated models, particularly Transformers, 
tend to make predictions in a more straightforward manner.
To substantiate this argument, 
we examine three variants whose accessible sequential information is progressively disrupted 
by different shuffling strategies:
\begin{align*}
  \begin{array}{rl}
  \text{\ding{172} Chronologically ordered:} & \mathbb{P}\left(v_{t+1}\,|\,v_t, v_{t-1}, \ldots, v_1; \Theta \right), \\
  \text{\ding{173} Partially shuffled:} & \mathbb{P}\left(v_{t+1}\,|\, v_t, \{v_1, v_2, \ldots \}; \Theta \right), \\
  \text{\ding{174} Completely shuffled:} & \mathbb{P}\left(v_{t+1}\,|\,\{v_1, v_2, \ldots, v_t\}; \Theta \right).
  \end{array}
\end{align*}
Here, $\{\cdot\}$ denotes a permuted set (implemented through a random shuffle operation).
The first keeps the original chronological order, 
thereby allowing the model to capture all sequential patterns.  
The second is partially shuffled except for the last item; consequently, 
the remaining interactions primarily serve as auxiliary cues for inferring the user's general identity.  
The third strategy applies a completely random shuffle, 
indeed degenerating into non-sequential prediction.
It is noteworthy that the three variants 
are trained under the same conventional protocol (\ie, sharing the same $\Theta$),
differing only in the input sequence order during inference.

We examine in \figurename~\ref{fig-order} how NDCG@10 changes as the sequence order is progressively disrupted. 
To ensure the reliability of this evaluation, 
\figurename~\ref{fig-order} presents results across three classical models and seven publicly available datasets spanning diverse domains and platforms, 
while accounting for variations in embedding sizes, training objectives, and dataset filtering strategies.
We have the following key observations:
\begin{itemize}[leftmargin=*]
  \item \ding{172} $\approx$ \ding{173}:
  Unlike the fully chronological order, which allows for the exploitation of all potential sequential patterns, 
  in the partially shuffled variant only the most recently interacted item $x_t$ preserves minimal temporal relevance \wrt~the target item.
  If advanced sequential recommenders truly rely on sequential information for complex yet more accurate inference, 
  such shuffling would be expected to cause an irreversible performance degradation.
  However, although SASRec and HSTU are more expressive than GRU4Rec and thus achieve greater performance gains, 
  both models appear rather insensitive to this disruption.
  Moreover, even the introduction of time-based positional encoding (HSTU) fails to alter this situation.
  \item \ding{172}/\ding{173} $\gtrapprox$ \ding{174}:
  Randomly shuffling the entire sequence completely disrupts the involved sequential information.
  Consequently, these `sequential' models typically fail to achieve competitive performance under such conditions. 
  This suggests that in most practical scenarios, at least the latest interaction still plays an indispensable role
  for next-item prediction.
\end{itemize}
The observations above collectively signify a somewhat counterintuitive conclusion: 
state-of-the-art sequential recommenders do not perform complex inference to make use of sequential dependencies.
Instead, they primarily summarize general user identify while emphasizing the latest interaction.
Formally,
\begin{align}
  \mathbb{P}\left(
    v_{t+1}\,|\,v_{t}, v_{t-1}, \ldots, v_1; \Theta
  \right)
  \approx \mathbb{P}(
    v_{t+1}\,|\,v_{t}, \underbrace{\{v_1,  v_2, \ldots\}}_{\text{non-sequential}}; \Theta
  ).
\end{align}
This suggests the `Markovian' nature of next-item recommendation:
conditioning on the entire interaction sequence is essentially equivalent to conditioning on the most recent interaction 
along with a non-sequential component.

\begin{remark}
  It is worth noting that the `Markovian' nature of next-item recommendation 
  is closely related to the notions of short-term and long-term user interests~\cite{yu2019adaptive,an2019neural,xiao2024joint,zheng2022disentangling},
  and has been exploited in the design of specialized architectures and learning paradigms.
  For instance, STAMP~\cite{liu2018stamp} explicitly processes the most recently interacted item to capture short-term interests, 
  while CL4Rec~\cite{xie2022contrastive} performs random shuffling to encourage the model to rely less on the input sequence order.
  In this work, we further reveal that the so-called short-term interest probably resides in the most recently interacted item, 
  and the long-term interest appears to be largely non-sequential.
\end{remark}

\begin{figure*}
	\centering
	\includegraphics[width=0.97\textwidth]{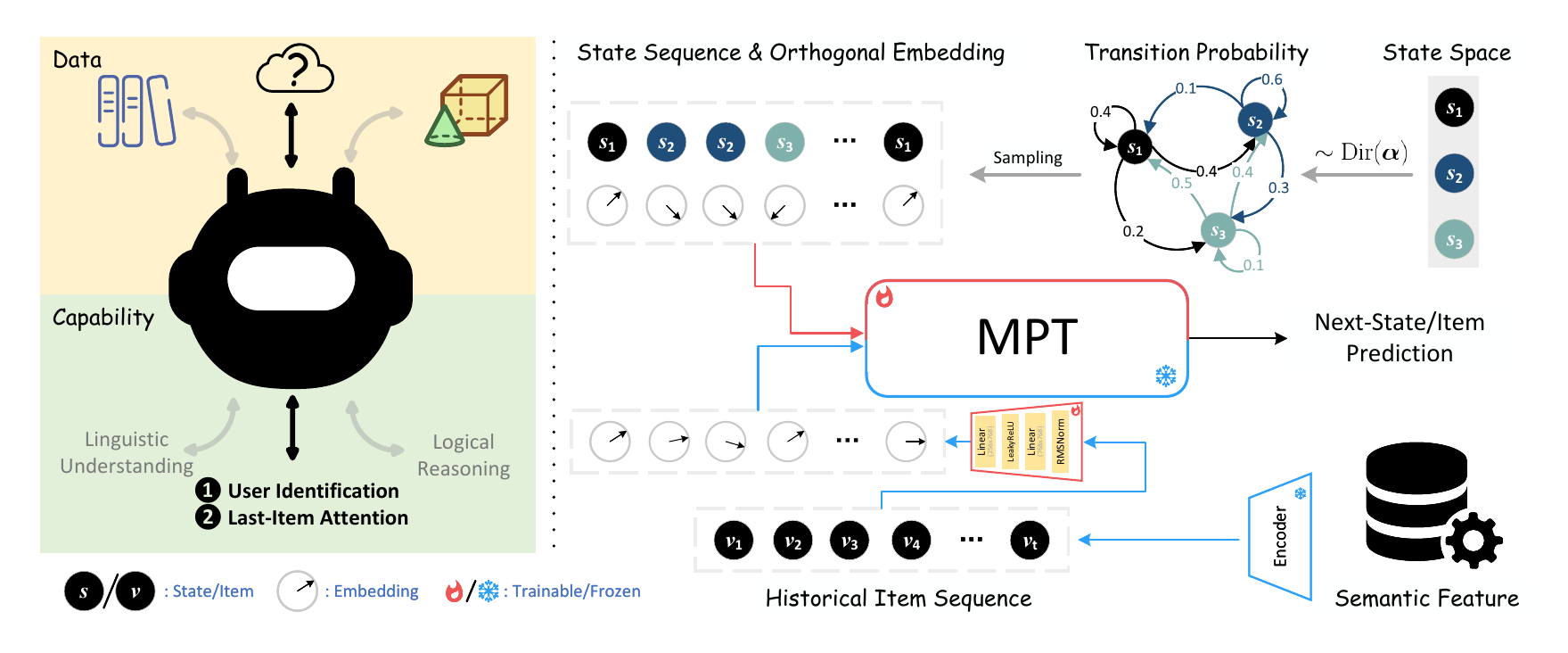}
    \caption{
      Overview of MPT.
      \textbf{Left:} Analogous to how linguistic corpora and mathematical data stimulate language understanding and logical reasoning, 
      we aim to identify what types of data can effectively activate the capabilities necessary for next-item recommendation.
      \textbf{Right:} Pipeline of Markovian pre-training and recommendation fine-tuning.
    }
	\label{fig-framework}
\end{figure*}

\section{Markovian Pre-Trained Transformer}
\label{section-mpt}

It is infeasible to exhaustively conduct the above experiments across all possible models and scenarios;
nonetheless, 
we believe these studies have already revealed the transferable capabilities necessary 
for developing specialized pre-trained recommenders:
\ding{182} effectively summarizing users' general identify based on historical interactions;  
\ding{183} placing particular emphasis on the most recently interacted item.

The remaining challenge lies in how to effectively stimulate these capabilities. 
Public recommendation datasets, which have already been extensively explored in prior studies, 
might not be a feasible choice in this context due to their heterogeneity ranging from diverse user behaviors and non-negligible domain gaps. 
Motivated by the fundamental role of mathematical data in enhancing the logical reasoning ability of LLMs, 
we turn to synthetic Markov chains.
Recall that a Markov chain is a discrete stochastic process in which the next possible state, $s_{t+1}$, is determined solely by the current state, $s_{t}$;
that is,
\begin{align*}
  \mathbb{P}(s_{t+1}\,|\,s_t, s_{t-1}, \ldots, s_1) = \mathbb{P}(s_{t+1}\,|\,s_t), \\
  s_{n} \in \mathcal{S}, \quad \forall n=1,2,\ldots, t+1,
\end{align*}
where $\mathcal{S}$ denotes the state space.
To predict the most likely next state in a sampled trajectory \textit{without any prior},
one must (I) identify the current state $s_t \in \mathcal{S}$ and (II) estimate the transition probabilities based on the observed frequencies 
from $s_t$ to the other states.
If a sequential model is capable of next-state prediction solely based on the context rather than memorization,
it must possess at least the following two capabilities:  
\ding{182} adaptively summarizing the transition probabilities from the sampled trajectory alone;
\ding{183} placing particular attention on the current state.
These precisely correspond to the required recommendation capabilities.
Fortunately, the Transformer architecture has been shown to exhibit such in-context learning capabilities in this regard~\cite{lepage2025markov}.
Thus, it is reasonable to expect that MPT could serve as a universal pre-trained model for next-item recommendation.
In the following, we provide a detailed introduction to the Markovian pre-training and recommendation fine-tuning paradigm.

\subsection{Markovian Pre-Training}

As illustrated in \figurename~\ref{fig-framework},  
MPT is fully pre-trained on synthetic Markov chains via a next-state prediction task:
\begin{align}
  \mathcal{L}_{\text{NSP}}(\Theta) = 
  \underset{\mathbf{P} \sim \text{Dir}(\bm{\alpha})}{\mathbb{E}} 
  \underset{\{s_t\}_{t=2}^T \sim \mathbf{P}|s_1}{\mathbb{E}}
  -\sum_{t=1}^{T-1} \log \mathbb{P}(s_{t+1}\,|\,s_{t}, \ldots, s_{1}; \Theta).
\end{align}
At each training step, a batch of trajectories is independently sampled based on the transition probability matrices $\mathbf{P}$,
which are generated from a Dirichlet prior $\text{Dir}(\bm{\alpha})$.
There are some key components affecting the behaviors of the Markov chains:
\begin{itemize}[leftmargin=*]
  \item \textbf{State space size $|\mathcal{S}|$}.
  Note that accurately estimating transition probabilities requires a sufficient number of observed transitions for each state $s \in \mathcal{S}$.
  Given the fixed maximum input length of a Transformer (\eg, $T = 1024$ in this paper), 
  it is necessary to control the number of states to ensure that most predictions remain meaningful and tractable;
  otherwise, MPT will just learn to make wild guesses.
  \item \textbf{Dirichlet distribution $\text{Dir}(\bm{\alpha})$}.
  The concentration vector $\bm{\alpha} \in \mathbb{R}_+^{|\mathcal{S}|}$ controls the uniformity of the transition probabilities.  
  For simplicity, this vector is assigned the identical value of $\alpha$ across all components.
  As $\alpha$ increases, the current state $s_t$ is more likely to transition to other states with roughly equal probability; 
  conversely, a smaller $\alpha$ causes some states to be revisited frequently while others are seldom visited.
  For next-item recommendation, we find that pre-training MPT with a slightly smaller $\alpha$ (\eg, 0.05) yields better results.  
  On the one hand, considering the $|\mathcal{S}|^2$ possible combinations, 
  the estimation of a `\textit{dense}' transition matrix becomes an over-parameterization problem if every combination carries a non-negligible probability.
  On the other hand, item-to-item relationships in practical scenarios tend to be highly sparse.
\end{itemize}
Recall that MPT in this study is expected to serve as a universal model for various downstream recommendation scenarios. 
Therefore, we promote insensitivity to the specific input space 
by generating orthogonal vectors as the state representations, rather than relying on a conventional learnable embedding table.
This operation is also performed independently for each sampling trajectory, ensuring that MPT never encounters the same input twice.

\textbf{Scaling laws}, which describe how loss scales as a power law \wrt~data volume and model size, 
have been extensively investigated in both large language models~\cite{kaplan2020scaling, hoffmann2022scaling} and recommender systems~\cite{zhai2024hstu, zhang2024wukong}. 
However, the scaling behavior has an inherent limit in Markov pre-training.
Given a Dirichlet prior $\text{Dir}(\alpha)$, the posterior transition probabilities also follow a Dirichlet distribution~\cite{lin2016dirichlet} improved by the observed occurrences  (see Appendix~\ref{section-theoretical-limit} for details).
Consequently, Markovian pre-training admits a well-defined Bayes estimator~\cite{berger2013statistical},
also implying that the achievable performance is inherently governed by this theoretical limit. 
As shown in Section~\ref{section-scaling}, MPT with four layers and a hidden size of 256 nearly reaches this limit. 
Further increasing the model capacity yields marginal gains and is therefore not cost-effective.

\begin{table}[]
  \caption{Dataset statistics}
  \label{table-dataset}
\begin{tabular}{crrrr}
  \toprule
Dataset       & \multicolumn{1}{c}{\#Users} & \multicolumn{1}{c}{\#Items} & \multicolumn{1}{c}{\#Interactions} & \multicolumn{1}{c}{Avg. Len.} \\
  \midrule
Beauty        & 22,363                      & 12,101                      & 198,502                            & 8.88                          \\
Toys          & 19,412                      & 11,924                      & 167,597                            & 8.63                          \\
Sports        & 35,598                      & 18,357                      & 296,337                            & 8.32                          \\
Yelp          & 77,277                      & 45,638                      & 2,103,896                          & 27.23                         \\
Online Retail & 16,520                      & 3,469                       & 519,906                            & 31.47                         \\
  \bottomrule
\end{tabular}
\end{table}

\begin{table*}[]
  \caption{
    Overall performance comparison.
    The best results of baselines and ours are separately marked in \textbf{bold} .
    `$\dagger$' indicates ID-based models with learnable ID embeddings.
    `$\Delta_1$': relative performance gap between MPT and the best baseline trained from scratch.
    `$\Delta_2$': relative performance gap between MPT and the best pre-trained baseline.
  }
  \label{table-overall-comparison}
\setlength{\tabcolsep}{3.5pt}
  \scalebox{0.86}{
\begin{tabular}{cl|ccccc|cccc|cc|rr}
  \toprule
\multicolumn{1}{l}{}                    &         & \multicolumn{5}{c}{Training from scratch}                              & \multicolumn{4}{c}{Pre-training \& Fine-tuning}         & \multicolumn{2}{c}{MPT}           &  &  \\
\multicolumn{1}{l}{}                    &         & GRU4Rec$^{\dagger}$         & SASRec$^{\dagger}$ & FMLPRec$^{\dagger}$ & HSTU$^{\dagger}$            & SASRec+$^{\dagger}$         & UniSRec & RecFormer & E4SRec$^{\dagger}$          & Qwen2.5         & Adaptor         & +LoRA           &     \multicolumn{1}{c}{$\Delta_1$}               &     \multicolumn{1}{c}{$\Delta_2$}               \\
  \midrule
\multirow{5}{*}{\rotatebox{90}{\textbf{Beauty}}}        & HR@1    & 0.0098          & 0.0161 & 0.0206  & 0.0251          & \textbf{0.0293} & 0.0073  & 0.0086    & \textbf{0.0236} & 0.0194          & \textbf{0.0329} & 0.0320          & 12.3\%             & 39.3\%             \\
                                        & HR@10   & 0.0524          & 0.0753 & 0.0775  & 0.0902          & \textbf{0.0982} & 0.0803  & 0.0676    & 0.0773          & \textbf{0.0826} & \textbf{0.1118} & 0.1100          & 13.9\%             & 35.4\%             \\
                                        & HR@20   & 0.0788          & 0.1060 & 0.1072  & 0.1246          & \textbf{0.1348} & 0.1233  & 0.1017    & 0.1046          & \textbf{0.1200} & \textbf{0.1531} & 0.1514          & 13.6\%             & 24.2\%             \\
                                        & NDCG@10 & 0.0278          & 0.0417 & 0.0454  & 0.0535          & \textbf{0.0591} & 0.0382  & 0.0333    & \textbf{0.0469} & 0.0464          & \textbf{0.0673} & 0.0658          & 13.8\%             & 43.3\%             \\
                                        & NDCG@20 & 0.0344          & 0.0495 & 0.0529  & 0.0621          & \textbf{0.0683} & 0.0490  & 0.0419    & 0.0538          & \textbf{0.0558} & \textbf{0.0777} & 0.0763          & 13.7\%             & 39.1\%             \\
  \midrule
\multirow{5}{*}{\rotatebox{90}{\textbf{Toys}}}          & HR@1    & 0.0126          & 0.0215 & 0.0254  & 0.0320          & \textbf{0.0334} & 0.0055  & 0.0087    & \textbf{0.0250} & 0.0175          & \textbf{0.0322} & 0.0316          & -3.8\%             & 28.8\%             \\
                                        & HR@10   & 0.0490          & 0.0846 & 0.0855  & 0.0981          & \textbf{0.1040} & 0.0829  & 0.0842    & \textbf{0.0798} & 0.0784          & \textbf{0.1202} & 0.1192          & 15.5\%             & 42.7\%             \\
                                        & HR@20   & 0.0704          & 0.1115 & 0.1124  & 0.1288          & \textbf{0.1393} & 0.1235  & 0.1221    & 0.1061          & \textbf{0.1114} & \textbf{0.1617} & 0.1611          & 16.1\%             & 31.0\%             \\
                                        & NDCG@10 & 0.0282          & 0.0493 & 0.0521  & 0.0610          & \textbf{0.0645} & 0.0380  & 0.0405    & \textbf{0.0492} & 0.0433          & \textbf{0.0705} & 0.0701          & 9.4\%              & 43.3\%             \\
                                        & NDCG@20 & 0.0335          & 0.0561 & 0.0589  & 0.0687          & \textbf{0.0733} & 0.0483  & 0.0501    & \textbf{0.0558} & 0.0515          & \textbf{0.0809} & 0.0806          & 10.4\%             & 45.0\%             \\
  \midrule
\multirow{5}{*}{\rotatebox{90}{\textbf{Sports}}}        & HR@1    & 0.0046          & 0.0068 & 0.0105  & 0.0139          & \textbf{0.0156} & 0.0056  & 0.0042    & 0.0092          & \textbf{0.0095} & \textbf{0.0162} & 0.0155          & 4.3\%              & 70.4\%             \\
                                        & HR@10   & 0.0259          & 0.0405 & 0.0413  & 0.0550          & \textbf{0.0576} & 0.0446  & 0.0345    & 0.0387          & \textbf{0.0473} & \textbf{0.0651} & 0.0644          & 12.9\%             & 37.7\%             \\
                                        & HR@20   & 0.0407          & 0.0607 & 0.0593  & 0.0788          & \textbf{0.0824} & 0.0699  & 0.0555    & 0.0563          & \textbf{0.0712} & 0.0924          & \textbf{0.0927} & 12.5\%             & 30.2\%             \\
                                        & NDCG@10 & 0.0135          & 0.0212 & 0.0238  & 0.0317          & \textbf{0.0337} & 0.0221  & 0.0167    & 0.0218          & \textbf{0.0256} & \textbf{0.0373} & 0.0365          & 10.8\%             & 45.7\%             \\
                                        & NDCG@20 & 0.0172          & 0.0263 & 0.0283  & 0.0377          & \textbf{0.0399} & 0.0285  & 0.0220    & 0.0262          & \textbf{0.0316} & \textbf{0.0442} & 0.0436          & 10.7\%             & 39.7\%             \\
  \midrule
\multirow{5}{*}{\rotatebox{90}{\textbf{Yelp}}}          & HR@1    & 0.0028          & 0.0001 & 0.0005  & \textbf{0.0051} & 0.0031          & 0.0050  & 0.0034    & 0.0035          & \textbf{0.0060} & 0.0061          & \textbf{0.0064} & 26.3\%             & 8.0\%              \\
                                        & HR@10   & 0.0246          & 0.0298 & 0.0334  & 0.0453          & \textbf{0.0466} & 0.0343  & 0.0247    & 0.0291          & \textbf{0.0411} & 0.0461          & \textbf{0.0495} & 6.3\%              & 20.5\%             \\
                                        & HR@20   & 0.0448          & 0.0539 & 0.0606  & 0.0773          & \textbf{0.0796} & 0.0567  & 0.0429    & 0.0513          & \textbf{0.0682} & 0.0783          & \textbf{0.0832} & 4.4\%              & 21.9\%             \\
                                        & NDCG@10 & 0.0116          & 0.0123 & 0.0138  & \textbf{0.0215} & 0.0209          & 0.0170  & 0.0121    & 0.0139          & \textbf{0.0204} & 0.0224          & \textbf{0.0241} & 12.0\%             & 17.8\%             \\
                                        & NDCG@20 & 0.0166          & 0.0184 & 0.0207  & \textbf{0.0295} & 0.0292          & 0.0226  & 0.0166    & 0.0195          & \textbf{0.0272} & 0.0305          & \textbf{0.0325} & 10.1\%             & 19.6\%             \\
  \midrule
\multirow{5}{*}{\rotatebox{90}{\textbf{Online Retail}}} & HR@1    & \textbf{0.0139} & 0.0014 & 0.0029  & 0.0092          & 0.0128          & 0.0135  & 0.0177    & \textbf{0.0205} & 0.0196          & 0.0159          & \textbf{0.0179} & 28.4\%             & -12.5\%            \\
                                        & HR@10   & 0.1175          & 0.1245 & 0.1333  & 0.1452          & \textbf{0.1589} & 0.1424  & 0.1407    & \textbf{0.1551} & 0.1503          & \textbf{0.1608} & 0.1568          & 1.2\%              & 3.7\%              \\
                                        & HR@20   & 0.1931          & 0.2014 & 0.2237  & 0.2304          & \textbf{0.2575} & 0.2275  & 0.2194    & \textbf{0.2528} & 0.2356          & \textbf{0.2572} & 0.2502          & -0.1\%             & 1.8\%              \\
                                        & NDCG@10 & 0.0561          & 0.0524 & 0.0548  & 0.0656          & \textbf{0.0719} & 0.0662  & 0.0683    & \textbf{0.0749} & 0.0731          & 0.0741          & \textbf{0.0746} & 3.8\%              & -0.4\%             \\
                                        & NDCG@20 & 0.0751          & 0.0718 & 0.0775  & 0.0870          & \textbf{0.0968} & 0.0877  & 0.0881    & \textbf{0.0995} & 0.0945          & \textbf{0.0982} & 0.0980          & 1.5\%              & -1.3\%             \\
  \bottomrule
\end{tabular}
  }
\end{table*}

\subsection{Recommendation Fine-Tuning}

Although extensive Markovian pre-training endows the model with the required capabilities,
recommendation fine-tuning remains necessary due to the mismatch between the pre-training and the downstream task space.
In line with prior pre-trained recommenders, we use semantic features as the raw item representations. 
For instance, for the Amazon Reviews dataset, 
each item's \textit{title}, \textit{categories}, and \textit{brand} information are jointly encoded by a sentence transformer (\textit{sentence-T5-xl} in this study) to obtain a unified vector representation.
Remark that MPT is technically agnostic to textual, visual, and other forms of item representation.
Compared with conventional learnable item embeddings, 
semantic features have been shown to yield comparable performance 
while being particularly well suited for cold-start items~\cite{yuan2023morec}.
This is more consistent with the principles of transferability and universality pursued in this work.

As shown in \figurename~\ref{fig-framework}, 
the recommendation fine-tuning pipeline employs only a lightweight input adaptor, 
consisting of an RMSNorm~\cite{zhang2019root} followed by a two-layer MLP with a LeakyReLU activation.
One may explore other parameter-efficient tuning techniques (\eg, LoRA); 
however, it should be emphasized that the learnt capabilities of MPT might be disrupted once its inherent parameters are re-tuned.
This issue stems from the heterogeneity present in certain recommendation datasets, 
where adapting MPT to highly specific scenarios would sometimes degrade its general capabilities.
In other words, in most scenarios, we only need to focus on aligning the item space with MPT.
This not only significantly reduces the overhead of fine-tuning but also suggests the generalizability of the aforementioned capabilities.

MPT is fine-tuned using a next-item prediction objective:
\begin{align}
  \mathcal{L}_{\text{NIP}}(\theta) = -\sum_{t=1}^{T-1} \log \mathbb{P}(v_{t+1} \mid v_t, \ldots, v_1; \Theta, \theta).
\end{align}
Here, $\theta$ denotes the additional learnable parameters involved in the adaptor.
In particular, we employ cosine similarity rather than the inner product as the scoring function, 
and incorporate a temperature hyperparameter to achieve appropriate rescaling.
This is a common practice in sequential recommendation~\cite{zhai2024hstu,hou2022unisrec}.
Note that $\mathcal{L}_{\text{NIP}}$ intrinsically corresponds to a cross-entropy loss computed over the entire item corpus, 
which may raise memory concerns in practical deployments.  
Although memory-efficient alternatives~\cite{klenitskiy2023turning,xu2024llm} to cross-entropy exist, 
we retain this objective by default to avoid introducing additional engineering factors.

\section{Experiments}

In this section, we aim to comprehensively evaluate the universality and transferability of MPT 
and conduct an initial examination of its underlying inference mechanism.

\textbf{Datasets.}
We consider five public datasets originating from three distinct platforms. 
The first three, Beauty, Toys, and Sports, are derived from the Amazon Reviews data, 
each processed using a 5-core filter that removes users and items with fewer than five interactions. 
Yelp is a real-world dataset comprising user reviews and check-ins; in this work, we use the 2018 version with a 10-core filter. 
Online Retail contains transaction records from a UK-based e-commerce platform, and we directly adopt the preprocessed version provided by~\cite{hou2022unisrec}.
Note that the Tmall dataset is not considered here because it lacks textual features required by most pre-trained baselines.
Following prior work~\cite{kang2018sasrec,hou2022unisrec}, 
we adopt a \textit{leave-one-out} fashion for dataset partitioning, 
where the final interaction is reserved for testing and the second-to-last interaction is used for validation.
The dataset statistics are presented in Table~\ref{table-dataset}.

\textbf{Evaluation metrics.}
The predicted scores over \textit{all items} are first ranked in descending order to yield a top-$N$ user-wise candidate list.
We employ the widely used HR@$N$ (Hit Rate) and NDCG@$N$ (Normalized Discounted Cumulative Gain) to assess the recommendation quality. 
The former measures the rate of successful hits among the top-$N$ recommended candidates, 
while the latter further accounts for positional relevance by assigning higher weights to correctly predicted items that appear earlier in the ranking.

\textbf{Baselines.}
To evaluate the potential of MPT as a universal recommendation model,
we consider UniSRec~\cite{hou2022unisrec} and RecFormer~\cite{li2023recformer}
as the representative recommendation-oriented pre-trained models.
To further compare the differences of Markovian pre-training and language pre-training for next-item recommendation, 
we also include E4SRec~\cite{li2023e4srec} that fine-tunes an instruction-tuned Llama-2~\cite{touvron2023llama} backbone,
as well as Qwen-2.5-7B~\cite{qwen2.5} that is fine-tuned in the completely same manner as MPT via an input adaptor.
In addition, we include four traditional sequential models, GRU4Rec~\cite{hidasi2015gru4rec}, SASRec~\cite{kang2018sasrec}, FMLPRec~\cite{zhou2022fmlprec}, and HSTU~\cite{zhai2024hstu}, 
to facilitate a straightforward comparison between pre-trained models and those trained from scratch.
As suggested by~\cite{petrov2023gsasrec,klenitskiy2023turning,xu2024llm}, 
SASRec+, which is trained using cross-entropy over the entire item corpus, 
is also considered to eliminate unfairness arising from differences in loss objectives.

\textbf{Implementation details.}
We adopt a Llama-style architecture with four hidden layers as the backbone of MPT. 
The model is pre-trained using the AdamW~\cite{loshchilov2019adamw} optimizer with a fixed learning rate of 3e-4 and a weight decay of 0.1. 
By default, convergence is achieved after learning approximately 10B tokens, corresponding to 9,600,000 trajectories of length 1,024.
During the recommendation fine-tuning stage, 
all datasets use a unified learning rate of 0.001 and an attention dropout rate of 0.2. 
Except for the Online Retail dataset, which uses a slightly larger weight decay of 1 to prevent overfitting, all other datasets employ a weight decay of 0.1.
For a fair comparison, we fix the embedding size at 256 for MPT as well as for all traditional sequential models.
For other hyperparameters, please refer to \tablename~\ref{table-hyperparameter} in Appendix~\ref{section-hyperparameter}.

\subsection{Overall Comparison}

\begin{figure}[t]
	\centering
    \includegraphics[width=0.47\textwidth]{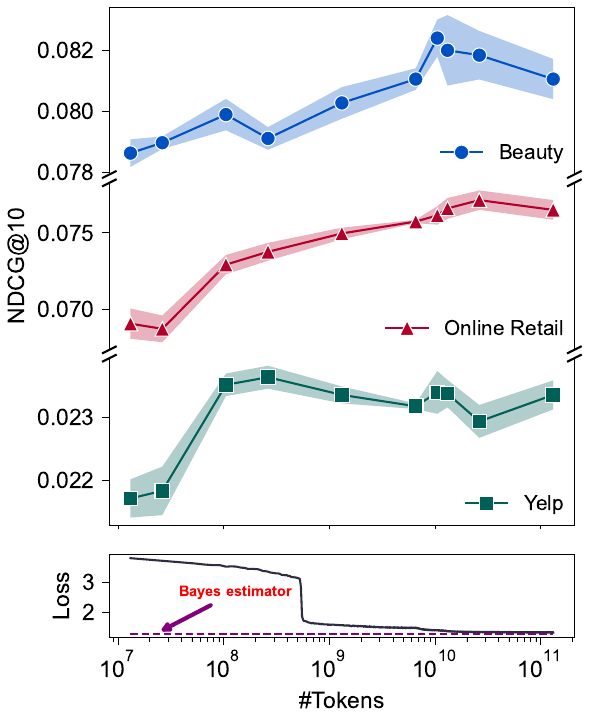}
    \caption{
      Data scaling: how the final recommendation performance evolves as more tokens are incorporated.
      Top: the validation NDCG@10 performance on the Beauty, Online Retail, and Yelp datasets.
      Bottom: the $\mathcal{L}_{\text{NSP}}$ loss, which reflects the accuracy of MPT in predicting the next possible state.
    }
	\label{fig-data-scaling}
\end{figure}

\begin{figure}[t]
	\centering
    \includegraphics[width=0.47\textwidth]{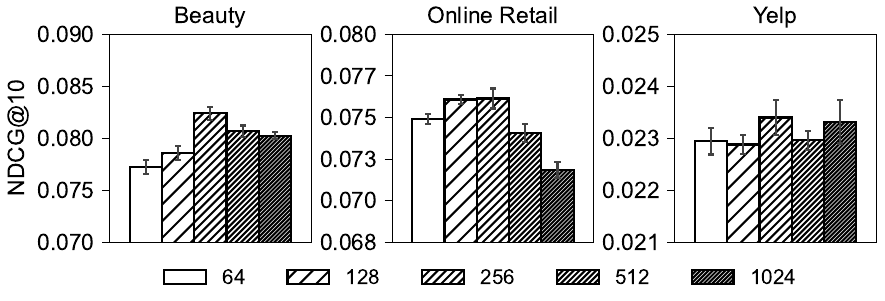}
    \caption{
      Model scaling: the validation NDCG@10 performance under varying hidden dimension sizes.
    }
	\label{fig-model-scaling}
\end{figure}

\begin{figure}[t]
	\centering
    \includegraphics[width=0.47\textwidth]{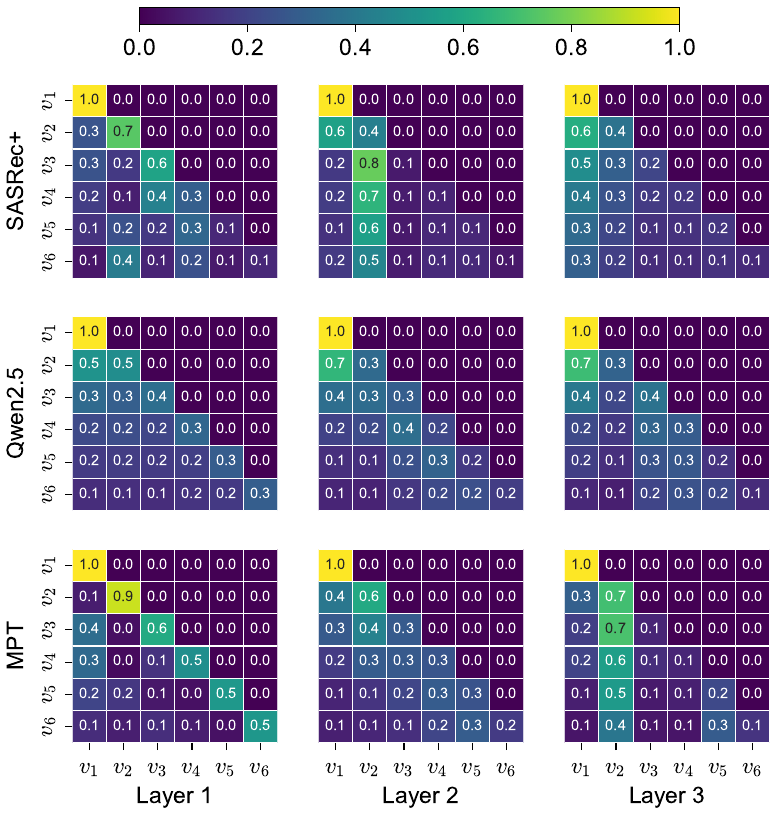}
    \caption{
      Attention maps of SASRec+, Qwen2.5, and MPT over a user sequence.
      The inference backbones of both Qwen2.5 and MPT are not fine-tuned using any recommendation data.
    }
	\label{fig-attention}
  \vspace{-1em}
\end{figure}

In this section, we investigate the universality of MPT 
by comparing its performance across five public datasets.
Remark that, the results of RecFormer, E4SRec, and Qwen2.5 are averaged over 3 independent runs in light of their substantial computational costs, 
whereas those of all other methods are averaged over 5 independent runs.
From Table~\ref{table-overall-comparison}, we derive the following key observations:
\begin{itemize}[leftmargin=*]
  \item 
  UniSRec and RecFormer are two recommendation-oriented approaches that have been pre-trained on several categories from Amazon Reviews.
  However, their performance is not only significantly inferior to that of specialized models trained from scratch, 
  but also rarely comparable to E4SRec and Qwen2.5 that are fully pre-trained on linguistic corpora,
  without incorporating any recommendation-specific information during pre-training.
  We hypothesize that, despite the adoption of sophisticated architectural designs and optimization strategies, 
  recommendation pre-trained models fail to learn universal and transferable capabilities due to the diverse user behaviors 
  and the non-negligible domain gaps.
  On the other hand, 
  although E4SRec and Qwen2.5 achieve competitive performance, such `successes' are not cost-effective 
  when their prohibitive inference overhead is taken into account (see Appendix~\ref{section-inference-time}).
  As discussed in Section~\ref{section-markovian-nature}, 
  previous efforts have been made without clearly identifying which capabilities a transferable recommender truly requires.
  \item
  The proposed MPT demonstrates outstanding performance by fine-tuning only a lightweight input adaptor.
  Even when compared with SASRec+, a specialized model trained using the cross-entropy loss, 
  MPT still achieves relative improvements of approximately 12\%.
  It is worth noting that, because Online Retail includes only item text for semantic representations, 
  MPT does not exhibit significant advantages over ID-based approaches.
  Furthermore, both MPT and Qwen2.5 are pre-trained without any recommendation-specific guidance and are fine-tuned in the same manner; however, Qwen2.5 underperforms MPT by about 32\%.
  Considering the substantial difference in model capacity that Qwen2.5 employs a hidden size of 3,584 while MPT uses only 256,
  the stronger capacity and thus more costs of Qwen2.5 does not translate into superior performance.
  We attribute this phenomenon to the more transferable capabilities acquired by MPT during the Markovian pre-training stage.
  This observation further suggests the superiority of Markovian pre-training over recommendation pre-training and language pre-training. 
  \item 
  Interestingly, 
  MPT with LoRA-based fine-tuning, in most cases, results in performance degradation compared with using an input adaptor alone.
  This phenomenon often arises when the pre-trained model's representations are already of sufficiently high quality.
  Similar behavior has been observed in other domains~\cite{zhai2022lit, tschannen2025siglip}; 
  however, to the best of our knowledge, this is the first report of such a phenomenon in the recommendation domain.
  Of course, the capabilities acquired through Markovian pre-training should not be expected to perfectly adapt to all scenarios.
  Indeed, LoRA-based fine-tuning on the Yelp dataset yields notable performance gains, 
  indicating that a more elaborate sequence summarization mechanism is required in such a case.
  We leave this direction for future work.
\end{itemize}

The following experiments are conducted under the input-adaptor-only setting, as it has been shown to be sufficiently effective.

\begin{figure*}
	\centering
  \subfloat[State space size $|\mathcal{S}|$]{
    \includegraphics[width=0.47\textwidth]{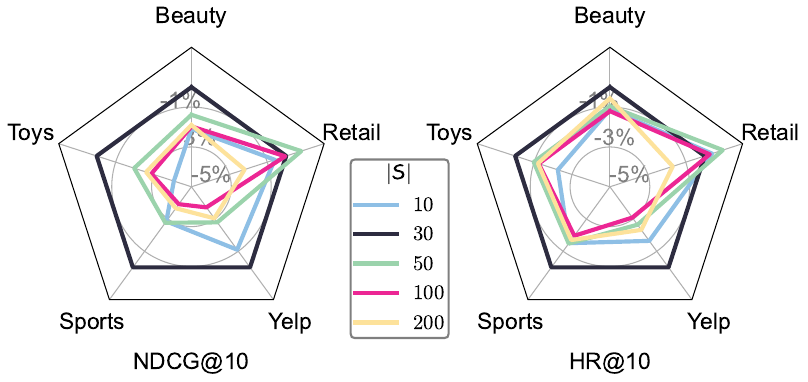}
    \label{fig-sensitivity-states}
  }
  \subfloat[ $\alpha$ of Dirichlet distribution]{
    \includegraphics[width=0.47\textwidth]{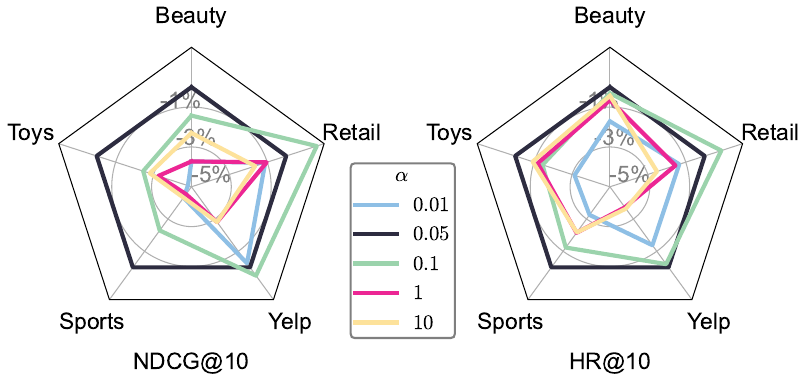}
    \label{fig-sensitivity-alpha}
  }
    \caption{
      Sensitivity analysis of state space size $|\mathcal{S}|$ and $\alpha$ of Dirichlet distribution.
      For clarity, we report the performance gaps relative to the default setting (\ie, $|\mathcal{S}| = 30$ and $\alpha = 0.05$).
    }
	\label{fig-sensitivity}
\end{figure*}

\subsection{Data and Model Scaling}
\label{section-scaling}

We illustrate in \figurename~\ref{fig-data-scaling} and \figurename~\ref{fig-model-scaling} how scaling the data volume and model size influences downstream recommendation performance.
Recall that, owing to the existence of a Bayes-optimal estimator, the performance of Markovian pre-training is theoretically upper-bounded.
The overall performance increases progressively until reaching a peak, after which it either stabilizes or declines slightly.
Although the number of tokens required to obtain the optimal performance varies across scenarios,
typically 10B tokens (corresponding to 9,600,000 trajectories of length 1,024) with a hidden dimension size of $256$ generally yield satisfactory results for all cases.
In addition, the loss dynamics differ markedly from those of conventional recommendation objectives. 
Several sudden drops can be observed in the loss curve. 
This phenomenon has also been reported in prior work~\cite{gopalani2024abrupt}, 
signifying the emergence of specialized capabilities.
Empirically, the checkpoint saved immediately after such a drop typically yields the best performance.

\subsection{Comparison of Inference Mechanisms}

In Section~\ref{section-mpt}, we have qualitatively elucidated why the capabilities acquired through Markovian pre-training 
may be transferable to next-item recommendation.
Here, we further corroborate  this argument by comparing the attention maps of SASRec+, Qwen2.5, and MPT.
As shown in \figurename~\ref{fig-attention}, both SASRec+ and MPT exhibit clear variations in attention patterns from Layer~1 to Layer~3.
In particular, MPT tends emphasize the most recently interacted items in the first layer, where an almost diagonal attention pattern can be observed.
This behavior is also consistent with the expected capability~\ding{183}.
In contrast, 
while the backbones of both Qwen2.5 and MPT are not fine-tuned using any recommendation data, 
the attention maps of Qwen2.5 unfortunately exhibit only imperceptible variations across Layers~1 to~3.
Consequently, Qwen2.5 fails to perform effective inference to capture users' general preferences, leading to inferior recommendation performance compared with SASRec+ and MPT.
This case study verifies the superiority of Markovian pre-training over language pre-training for next-item recommendation.

\subsection{Sensitivity Analysis}

Markovian pre-training involves two primary hyperparameters that affect the behaviors of the Markov chains: 
the state space size $|\mathcal{S}|$ and the concentration parameter $\alpha$ of the prior Dirichlet distribution. 
Both directly determine the degree of over-parameterization. 
Specifically, excessively large values result in an overly challenging task under the maximum input length of 1,024, 
whereas overly small values render the Markovian pre-training task too trivial to stimulate the required capabilities.
Consequently, as illustrated in \figurename~\ref{fig-sensitivity}, 
optimal performance in most scenarios is achieved around the default setting of $|\mathcal{S}| = 30$ and $\alpha = 0.05$.

\section{Related Work}

\textbf{Recommender systems} are designed to alleviate the rapidly growing problem of information overload and have become an indispensable component of modern applications such as e-commerce platforms~\cite{zhou2018din}, online news services~\cite{gong2022news}, and social media~\cite{sharma2024social}.
Next-item recommendation aims to predict a user's next likely interaction based on their historical sequence, 
which is generally assumed to reflect the user's underlying interests. 
Owing to the sequential nature of this task, prior studies have largely focused on exploiting various sequential architectures, 
ranging from RNNs~\cite{hidasi2015gru4rec} to Transformers~\cite{kang2018sasrec,sun2019bert4rec}.
The theories of short-term and long-term interest modeling~\cite{yu2019adaptive,an2019neural,xiao2024joint,zheng2022disentangling} 
have been widely applied to driving the advancement of architectural designs and optimization strategies.
The findings of this study may offer an alternative interpretation:
such approaches are effective probably because they explicitly emphasize the most recent interactions (as in STAMP~\cite{liu2018stamp}), 
while modeling long-term user interests in a non-sequential manner (as exemplified by CL4Rec~\cite{xie2022contrastive}).

\textbf{Pre-training and fine-tuning} 
has become the \textit{de facto} paradigm across a wide range of domains~\cite{chen2020simclr,devlin2019bert,radford2021clip}.
Pre-training endows models with strong generalization capabilities by exposing them to diverse scenarios, 
thereby enabling efficient transfer to downstream tasks. 
For instance, self-supervised visual representation learning~\cite{he2020moco,tschannen2025siglip} on large-scale image corpora can be effectively transferred to classification, detection, and segmentation tasks with only few-shot fine-tuning. 
Consequently, once effective pre-training for recommendation is achieved, repetitive and costly retraining can be largely avoided, 
and changes in community preferences can be rapidly accommodated through lightweight fine-tuning.
Considerable efforts have been devoted to developing universal pre-trained recommender systems. 
ZESRec~\cite{ding2021zesrec}, a hierarchical Bayesian model with a probabilistic encoder-decoder architecture, 
explores the feasibility of zero-shot recommendation. 
UniSRec~\cite{hou2022unisrec} learns universal sequential representations via a MoE-enhanced adaptor together with effective sequence-item and sequence-sequence contrastive tasks. 
RecFormer~\cite{li2023recformer} turns to the textual space to enable transferable recommendation modeling.
More recently, some studies have turned to large language models (LLMs)~\cite{radford2018gpt,touvron2023llama} 
that are purely pre-trained on textual data.
Although LLMs do not explicitly incorporate recommendation-oriented patterns during pre-training, their strong linguistic understanding and logical reasoning capabilities are believed to hold significant potential for universal recommendation~\cite{liao2024llara,he2025llm2rec,wang2025frellm4rec}.
For example, E4SRec~\cite{li2023e4srec} achieves competitive performance by connecting item ID embeddings to LLaMA-2 through a linear adaptor combined with LoRA-based fine-tuning.

Note that recommendation pre-training is also closely related to multi-task~\cite{geng2022p5,wang2023multi} 
and cross-domain recommendation~\cite{tang2012cross,wu2025id}.
The former emphasizes universality in addressing diverse recommendation tasks (\eg, rating prediction and review summarization), 
rather than being limited to next-item recommendation. 
The latter focuses on how cross-domain learning benefits overlapping users and items. 
In contrast, we specifically study how a pre-trained recommender can be readily adapted to 
various next-item recommendation scenarios, without requiring overlapping users or items.

\section{Conclusion and Furture Work}

We analyze the inference mechanisms of advanced sequential recommenders and investigate the feasibility of using Markovian pre-training to stimulate the required capabilities.
To the best of our knowledge, 
this is the first time that fully synthetic data have been used to pre-train a recommendation model 
and achieves state-of-the-art performance via a lightweight fine-tuning. 
Compared with traditional recommendation pre-training and language pre-training, 
we demonstrate clear advantages of Markovian pre-training in both performance and efficiency. 
These findings suggest a promising new paradigm for building a universal recommender.

Markov chain data alone may not be sufficient to cover the full spectrum of recommendation scenarios.
This highlights the need to explore more sophisticated forms of synthetic data. 
For instance, Yin et al.~\cite{yin2024dataset} introduced a dataset regeneration technique aimed at enhancing cross-architecture generalizability. 
Additionally, recent research has utilized LLMs to simulate user behaviors~\cite{wang2025user,liu2025recoworld}, with the objective of generating more controllable and high-quality interaction data.
They collectively suggest a promising future for recommendation simulation~\cite{ie2019recsim,shi2019virtual}.

%%
%% The next two lines define the bibliography style to be used, and
%% the bibliography file.
\bibliographystyle{ACM-Reference-Format}
\bibliography{refs}

%%
%% If your work has an appendix, this is the place to put it.
\appendix

\section{Experimental Details}

\subsection{Baselines}
\label{section-baselines}

Here, we primarily introduce the involved baselines and explain how their results in \tablename~\ref{table-overall-comparison} were obtained.

\begin{itemize}[leftmargin=*]
\item \textbf{GRU4Rec}~\cite{hidasi2015gru4rec} was the first to apply an RNN architecture to session-based recommendation. 
It consists of a GRU~\cite{cho2014gru} kernel followed by a linear output projection layer.
\item \textbf{SASRec}~\cite{kang2018sasrec} is a classic Transformer-based recommender that employs a causal attention mechanism and learnable positional encoding. 
Following the official implementation, a binary cross-entropy loss is used by default.
\item \textbf{FMLPRec}~\cite{zhou2022fmlprec} is an MLP-based recommender capable of achieving competitive performance through the incorporation of learnable filters.
\item \textbf{HSTU}~\cite{zhai2024hstu} is known for its generative recommendation mechanism. 
Here, we employ the simplified variant (with learnable ID embeddings only) provided by the official implementation.
\item \textbf{SASRec+}~\cite{klenitskiy2023turning,xu2024llm} differs from SASRec in its training objective.
The cross-entropy loss instead of point-wise objectives is employed for superior performance.
\item \textbf{UniSRec}~\cite{hou2022unisrec} proposed a well-known universal recommender pre-trained across five categories from Amazon Reviews: `Grocery and Gourmet Food', `Home and Kitchen', `CDs and Vinyl', `Kindle Store', and `Movies and TV'. 
Consistent with MPT, we fine-tune only its MoE-enhanced adaptor for downstream tasks. 
\item \textbf{RecFormer}~\cite{li2023recformer} performs next-item recommendation based on item sentences. Unlike pure language pre-training, 
RecFormer incorporates recommendation-specific information in the  pre-training stage. 
It is pre-trained across seven categories from Amazon Reviews: `Automotive', `Cell Phones and Accessories', `Clothing, Shoes and Jewelry', `Electronics', `Grocery and Gourmet Food', `Home and Kitchen', and `Movies and TV'.
\item \textbf{E4SRec}~\cite{li2023e4srec} connects pre-trained item ID embeddings directly to an instruction-tuned LLM (Platypus2-7B~\cite{lee2023platypus} in this study) through a linear adaptor and LoRA-based fine-tuning.
\item \textbf{Qwen2.5-7B}~\cite{qwen2.5} is one of the most powerful open-source LLMs. We fine-tune it in the same manner as MPT to examine the differences between Markovian pre-training and language pre-training.
\end{itemize}

For traditional sequential recommendation methods, hyperparameters are tuned according to the validation results on the Beauty dataset. 
For other pre-trained methods, we adhere to the fine-tuning protocols specified in their respective original papers.

\subsection{Hyperparameters}
\label{section-hyperparameter}

\begin{table}[t]
  \caption{Hyperparameters of MPT during different stages.}
  \label{table-hyperparameter}
\begin{tabular}{c|c|ccccc}
  \toprule
                              &               & Beauty & Toys  & Sports & Yelp  & Retail \\
  \midrule
\multirow{7}{*}{\rotatebox{90}{Pre-training}} & Learning Rate & 0.001  & 0.001 & 0.001  & 0.001 & 0.001         \\
                              & Weight Decay  & 0.1    & 0.1   & 0.1    & 0.1   & 0.1           \\
                              & Max Len.             & 1024   & 1024  & 1024   & 1024  & 1024          \\
                              & $\alpha$         & 0.05   & 0.05  & 0.05   & 0.05  & 0.05          \\
                              & $|\mathcal{S}|$             & 30     & 30    & 30     & 30    & 30            \\
                              & \#Layers      & 4      & 4     & 4      & 4     & 4             \\
                              & \#Heads       & 2      & 2     & 2      & 2     & 2             \\
  \midrule
\multirow{5}{*}{\rotatebox{90}{Adaptor}}      & Learning Rate & 0.001  & 0.001 & 0.001  & 0.001 & 0.001         \\
                              & Weight Decay  & 0.1    & 0.1   & 0.1    & 0.1   & 1             \\
                              & Max Len.             & 50     & 50    & 50     & 50    & 50            \\
                              & Dropout Rate  & 0.2    & 0.2   & 0.2    & 0.2   & 0.2           \\
                              & Temperature  & 0.07    & 0.07   & 0.07    & 0.07   & 0.07           \\
  \midrule
\multirow{3}{*}{\rotatebox{90}{+LoRA}}         & LoRA\_Rank    & 16     & 16    & 16     & 16    & 16            \\
                              & LoRA\_Alpha   & 16     & 16    & 16     & 16    & 16            \\
                              & LoRA\_Dropout & 0.1    & 0.1   & 0.1    & 0.1   & 0.1           \\
  \bottomrule
\end{tabular}
\end{table}

We pre-training and fine-tuning MPT according to the hyperparameters presented in Table~\ref{table-hyperparameter}.
To accelerate the pre-training process, we adopt mixed-precision training in BF16 format.

\begin{figure*}[t]
	\centering
    \includegraphics[width=0.85\textwidth]{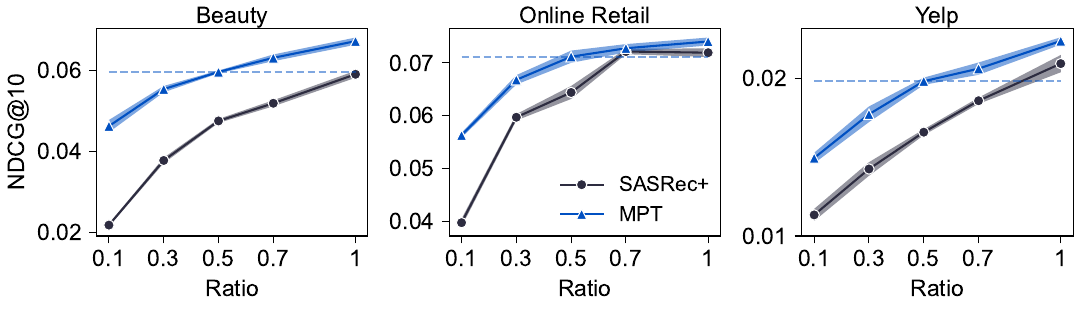}
  \caption{
    How recommendation performance evolves as an increasing number of user sequences are learnt during training.
    The dashed line indicates the performance of MPT trained with only half of the data, 
    which is nearly comparable to that of SASRec+ trained on the full dataset.
  }
	\label{fig-ratio}
\end{figure*}

\subsection{Inference Time}
\label{section-inference-time}

\begin{table}[t]
  \caption{
    Inference time. 
    The wall time (in seconds) is measured on an Intel(R) Xeon(R) Gold 6133 CPU paired with a single NVIDIA A800 GPU.
  }
  \label{table-inference-time}
\begin{tabular}{l|rrrrr}
  \toprule
          & Beauty & Toys   & Sports & Yelp    & Retail \\
  \midrule
GRU4Rec   & 1.09   & 1.11   & 1.55   & 3.82    & 1.10          \\
SASRec    & 1.20   & 1.04   & 1.69   & 3.62    & 0.90          \\
FMLPRec   & 1.88   & 1.60   & 2.80   & 6.86    & 1.49          \\
HSTU      & 1.92   & 1.63   & 2.97   & 7.71    & 1.95          \\
UniSRec   & 1.85   & 1.59   & 2.54   & 4.72    & 2.09          \\
Recformer & 351.99 & 270.24 & 590.25 & 1639.49 & 155.96        \\
E4SRec    & 251.06 & 274.17 & 397.60 & 1138.52 & 186.17        \\
Qwen2.5   & 299.56 & 158.53 & 332.80 & 759.32  & 151.74        \\
MPT       & 1.55   & 1.35   & 2.24   & 5.18    & 1.31          \\
  \bottomrule
\end{tabular}
\end{table}

As shown in Table~\ref{table-inference-time}, 
MPT achieves state-of-the-art performance while maintaining inference speed comparable to traditional models. 
In contrast, the latency of pre-trained language models is considerably higher, rendering them unsuitable for practical recommendation scenarios.

\section{Theoretical Limit of MPT}
\label{section-theoretical-limit}

A Markov chain can be regarded as a collection of $|\mathcal{S}|$ independent multinomial distributions $\{\text{Cat}(|\mathcal{S}|; \mathbf{p}_i)\}_{i=1}^{|\mathcal{S}|}$.
In this study, 
each ground-truth transition probability vector $\mathbf{p}_i$ is drawn from a Dirichlet prior $\text{Dir}(\bm{\alpha})$.
Because the Dirichlet distribution is conjugate to the multinomial distribution~\cite{lin2016dirichlet},
the posterior distribution of the transition probabilities remains Dirichlet. Specifically,
\begin{align*}
  \mathbf{p}_i \mid [s_1, s_2, \ldots, s_T] \sim \text{Dir}(c_{i,1} + \alpha_1,\, c_{i,2} + \alpha_2,\, \ldots,\, c_{i, |\mathcal{S}|} + \alpha_{|\mathcal{S}|}),
\end{align*}
where $c_{i,j}$ counts the number of transitions from state $i$ to state $j$ along the entire trajectory $[s_1, s_2, \ldots, s_{T}]$.
Hence, for an arbitrary trajectory, we can derive the Bayes-optimal estimator~\cite{berger2013statistical}
\begin{align*}
  \hat{p}_{ij} = \mathbb{E}\left[p_{ij}|[s_1, s_2, \ldots, s_T]\right] = \frac{c_{i,j} + \alpha_j}{\sum_{j=1}^{|\mathcal{S}|} (c_{i,j} + \alpha_j)}.
\end{align*}
This Bayes estimator indicates a theoretical upper bound that MPT can attain.

\section{Data Efficiency}

Pre-trained models are known for their data efficiency, capable of achieving comparable performance with significantly less data.
As shown in \figurename~\ref{fig-ratio}, MPT also exhibits this property.
However, it is important to emphasize that, in contrast to conventional pre-trained approaches, 
these benefits stem from transferable capabilities rather than from transferred knowledge.

\end{document}